\documentstyle[12pt,aaspp4,epsf]{article}
 
\begin{document}
\def\ltsim{ \,{}^<_\sim\, }
\def\gtsim{ \,{}^>_\sim\, }
\def\deg{\hbox{$^\circ$}}
\def\sq{\hbox{\rlap{$\sqcap$}$\sqcup$}}
\def\etal{{\it et al.}}
\def\eg{{\it e.g.}}
\def\ie{{\it i.e.}}
\def\cf{{\it cf.}}
\def\m{\vspace{0.18in}}

\title{UNVEILING PALOMAR 2: THE MOST OBSCURE GLOBULAR CLUSTER IN THE OUTER HALO}
\author{William E. Harris\altaffilmark{1}, Patrick R.~Durrell\altaffilmark{1},
Glen R. Petitpas, Tracy M. Webb, and Sean C. Woodworth}
\affil{Department of Physics and Astronomy, McMaster University \\
Hamilton, Ontario, L8S 4M1, Canada\\
Electronic mail:  harris,durrell,petitpa,webb,woodwrth@physics.mcmaster.ca}
 
\altaffiltext{1}{Visiting Astronomer, Canada-France-Hawaii Telescope, operated by 
the National Research Council of Canada, le Centre National de la Recherche 
Scientifique de France, and the University of Hawaii.}
 
\begin{abstract}
We present the first color-magnitude study for Palomar 2, a distant and
heavily obscured globular cluster near the Galactic anticenter. 
Our $(V,V-I)$ color-magnitude diagram (CMD), obtained with the 
UH8K camera at the CFHT, reaches $V(lim) \simeq 24$
and clearly shows the principal sequences of the cluster, though with
substantial overall foreground absorption and differential reddening.
The CMD morphology shows a well populated red horizontal branch with a
sparser extension to the blue, similar to clusters such as
NGC 1261, 1851, or 6229 with
metallicities near [Fe/H] $\simeq -1.3$.  From an average of several indicators,
we estimate the foreground reddening at $E(B-V) = 1.24 \pm 0.07$ and
obtain a true distance modulus $(m-M)_0 = 17.1 \pm 0.3$, placing it
$\simeq 34$ kpc from the Galactic center. 
We use starcounts of the bright stars to measure the core radius, 
half-mass radius, and central concentration of the cluster.  Its 
integrated luminosity is $M_V^t \simeq -7.9$, making it clearly 
brighter and more massive than most other clusters in the
outer halo.  Very rough arguments based on its half-mass radius and
radial velocity suggest that Palomar 2 
is now moving in toward perigalacticon 
on a highly elliptical orbit ($e \gtrsim 0.7$).

\end{abstract}

\section{Introduction}

Most globular clusters in the Milky Way are within a few kiloparsecs of the
Galactic center (\markcite{rac89}Racine \& Harris 1989), and it is common
knowledge that studies of these inner-halo objects 
are often badly hampered by large amounts
of reddening, differential reddening, and field-star contamination.
These inner clusters are now drawing the attention they deserve
(\eg\ \markcite{arm89}Armandroff 1989; 
\markcite{ort97}Ortolani \etal~1997; and references cited there).
It is something of a surprise, however, 
to realize that some of the rarer, remote
{\it outer}-halo clusters suffer similar afflictions which make them
quite difficult to study.  Of these, Palomar 2
($\alpha = 04^h 46^m 06^s, \delta = +31^o 22' 51''$ (J2000); 
$\ell = 170\fdg 5, ~b = -9\fdg 1$; see \markcite{har96}Harris 1996) 
is the most extreme example known.
Its coordinates give it the unique distinction of being the globular cluster
located farthest on the sky from the Galactic center.  Thanks to its moderately
low latitude, it is unfortunately sitting behind a thick wall of
absorption from the foreground Galactic disk, and it has not
attracted much previous attention from observers.  

The only previously published distance 
measurement for Palomar 2 is by \markcite{har80}Harris (1980), 
from a very rough estimate of the photographic magnitudes of the brightest giants 
in the cluster.  In this paper, we present new photometry which
reveals the nature of its color-magnitude diagram (CMD) for the first
time and permits several of its basic parameters (distance, reddening,
metallicity) to be determined with some reasonable level of confidence.  
In brief, we find that Palomar 2 does indeed belong to the outer halo, 
currently lying $\sim 34$ kpc from the Galactic center.
Since the remote halo clusters are essential to establishing 
the mass distribution of the Galaxy on its largest scales
(\eg\ \markcite{har78}Hartwick \& Sargent 1978;
\markcite{lit87}Little \& Tremaine 1987) and the evolutionary
history of the halo (\eg\ \markcite{ric96}Richer \etal~1996), 
each new addition to the short list of these objects is valuable. 

\section{Observations and Data Reduction}

The images for this study were obtained on 1996 September 19 -- 24
as a secondary project 
during an observing run on the Canada-France-Hawaii Telescope (CFHT).
We used the University of Hawaii 8K mosaic CCD camera with $V$ and $I$ filters.
This detector has an image scale of $0\farcs 21$ per pixel at prime focus;
for our exposures of Palomar 2 
we kept images from only ``chip 1'' (the best CCD
in the array of eight),  which was also the CCD used to record standard
stars throughout the run.  Each individual CCD in the mosaic has $2048 \times 4096$
pixels and thus a field size of $7\farcm 2 \times 14\farcm 3$.

For Palomar 2, we obtained three exposures in $V$
(each 300 seconds) and three in $I$ (60, 60, and 100 seconds).
The raw CCD frames were overscan- and bias-frame subtracted, then flat-fielded
to $\pm 1\%$ with master twilight flat exposures averaged over the entire run.
The seeing on all images was $0\farcs 5 - 0\farcs 6$, good enough to
resolve the bright stars easily all the way in to the cluster center.
A portion of our image field centered on Palomar 2 is shown in Figure~1.
Inspection of the frames suggested immediately that 
this cluster is moderately luminous, quite distant, 
but also heavily and patchily reddened.  This impression is strongly
confirmed by (for example) the POSS prints of the
large-scale area around Palomar 2, which show a large number of
patchy, dense dust clouds extending for many degrees over all parts
of the region.  On the blue POSS print in particular, the cluster
is only a faint, unresolved clump behind an obviously thick dust lane, 
in strong contrast to its appearance at redder wavelengths.

A master frame in each color was constructed by averaging
the three individual exposures.  Then to carry out the photometry, we used the suite of 
DAOPHOT II and ALLSTAR codes \markcite{ste92}(Stetson 1992) with a 
star-finding threshold set 4 standard deviations above the
sky noise on each frame.  The starlists from each frame were merged
by simple coordinate matching after registration to a common coordinate
system.  Finally, we rejected individual stars with internally uncertain
measurements ($\chi > 2$ in either
color).  The final list used to define the CMD contains 2607 stars (the catalog
of measurements is not reproduced here, but can be obtained electronically
on request to the first author).

The instrumental $(v,i)$ magnitude scales were calibrated against $(V,I)$ 
from images of standard-star fields from \markcite{lan92}Landolt (1992)
and \markcite{chr85}Christian \etal~(1985). The derived transformations are
\begin{eqnarray}
v & = & V + const + 0.15 X + 0.031(V-I) \\
i & = & I + const + 0.06 X + 0.007(V-I)
\end{eqnarray}
where $X$ is the airmass.  In these equations we use mean airmass coefficients
from \markcite{lan92}Landolt (1992), though our results are insensitive to
their exact values since the Palomar 2 frames were all taken near meridian transit
($X \simeq 1.02$).  The scatter of the standard stars about the mean
relations suggests that our zeropoints are accurate to $\pm0.02$ magnitude
in both $V$ and $I$.

The internal measurement errors of the photometry were evaluated through
artificial-star tests, with the standard tools in DAOPHOT.
We added artificial stars (scaled PSFs) to our images 
in groups of 200, inserting them into a region of 200 px radius ($40''$)
centered on the cluster (this central region
contains the vast majority of the
cluster members and is thus the most relevant area for this purpose.)
The resulting estimates of the internal rms errors $\sigma(V), \sigma(I)$
are displayed in Figure \ref{fig2}.
From the same artificial-star simulations, we also found that
the completeness of image detection is $>95\%$ for $V < 23.5$, but falls
to 70\% at $V=24$ and then steeply drops off at fainter levels.

\section{Data Analysis}

\subsection{CMD morphology}

The CMD for all measured stars is shown in Figure \ref{fig3}.  The presence
of a red-giant branch (RGB) with a bright tip at $V \sim 18.8, (V-I) \sim 3.2$
is evident, as is a clump of stars around $V \sim 21.5, (V-I) \sim 2.3$ which
we identify as a red horizontal branch (HB).  The flood of faint stars
starting near $V \sim 23.5$ is likely to be the 
subgiant branch and upper turnoff region of the cluster, 
though our data do not reach faint enough
to locate the unevolved main sequence; for a normal 
offset $\Delta V _{MSTO-HB} \simeq 3.5$ magnitudes, the turnoff point (MSTO)
would lie near $V \sim 24$.

Considerable scatter is present in all parts of the diagram, generated 
by a combination of (a) field-star contamination, (b) differential reddening,
and (c) random photometric errors.  
To minimize the field contamination, we experimented with CMD plots selected from different 
radial zones around the cluster center to find ones that isolated the cluster more clearly.
These trials showed that a good separation between the cluster stars and
the background field population could be obtained with a dividing line near 
$R \simeq 50''$.  Figure \ref{fig4} shows the CMDs for an inner zone ($R < 50''$)
and an outer one ($R > 120''$).  In the inner zone (left panel) 
the cluster sequences are now
more evident.  The scatter along the giant branch
and horizontal branch is now much less than in Fig.~3, but still about a
factor of two larger than expected from the random measurement errors
described above.
We therefore suggest (with additional evidence to be
discussed below) that differential reddening is responsible for much of
the residual scatter.

The sparsely populated outer zone is dominated 
by the miscellaneous scatter of field stars, except
for the faint clump at $V \simeq 24, (V-I) \simeq 2$.  
We suggest that many of these may be turnoff stars belonging to
the cluster;  interestingly, a high fraction of these faint 
stars are found preferentially
in the southeast quadrant of the field.  The spatial distribution of the faint
`blue' stars ($V > 23, (V-I) < 2.4$) is shown in Figure \ref{fig5}, while the
distribution of bright stars is shown in Figure \ref{fig6}
for comparison.  Careful inspection of the original images 
shows that there is indeed a preponderance of moderately faint stars to the
SE of cluster center, which do not show up in the same numbers
in the other quadrants.  Significant differential reddening appears to
be responsible for this rather unusual distribution.  
Because of their location in the CMD, 
the faint blue stars plotted in Fig. \ref{fig5} map out the distribution
of foreground reddening rather effectively:  since they are near the limiting
magnitude of the photometry, an extra 
half-magnitude of absorption (or $\Delta E(V-I) \sim 0.2$)
would be enough to push them below the detection limit, and there are no stars 
lying just above them in the CMD which would move down to replace them.

The CMD for the inner zone (Fig.~4a) clearly shows the red HB clump,
but also reveals a scattering of blue HB stars.  Comparison with the outer-zone
CMD (Fig.~4b) strongly suggests that these blue stars cannot be simply field
stars, since they occupy a part of the CMD that is nearly vacant in the outer zones.
To verify this statement, we generated
a cleaned version of the cluster CMD 
in which the field contamination was reduced to
its statistical minimum.  We defined an annulus $(100'' < R < 112'')$ 
taken from the outer zone with the same {\it area} as the inner $(R < 50'')$ circle.  
Then, for each star in this outer annulus, one star was
removed from the inner-circle CMD if any such star could be found at 
a magnitude and color within
a box of size $\Delta(V-I) = \pm0.1$ mag and $\Delta V = \pm1.0$ mag
centered on the outer star.
The results of this point-by-point removal 
are quite insensitive to the particular box size, since the inner
circle contains 1081 stars and the background annulus just 87 stars.

The resulting cleaned CMD is shown in Figure \ref{fig7}.
The blue HB component is still present, along with the more distinct red HB clump.
Morphologically, this type of HB distribution 
is relatively rare but not unprecedented.  Other
halo clusters which closely resemble it are NGC 1261 at [Fe/H] $=-1.35$ 
(\markcite{fer93}Ferraro \etal\ 1993), NGC 1851 at [Fe/H] $=-1.26$ 
(\markcite{wal92}Walker 1992), or NGC 6229 at [Fe/H] $=-1.44$
(\markcite{car91}Carney \etal\ 1991).  This type of
red HB morphology -- a heavily populated red HB with a sparser tail 
extending across well to the blue -- falls in a range where the HB distribution is
quite sensitive to metallicity.  For example,
M3 ([Fe/H] $= -1.57$) and M5 ([Fe/H] $= -1.33$) have
clearly bluer HBs (\eg\ \markcite{san82}Sandage \& Katem 1982; 
\markcite{san96}Sandquist \etal~1996); whereas
NGC 362 (\markcite{har82}Harris 1982), at [Fe/H] = $-1.16$, has a definitely
redder HB morphology.  (All quoted metallicity values are taken from 
the catalog of \markcite{har96}Harris 1996.)  From these comparisons, we suggest that
Pal 2 is likely to have a metallicity in the range [Fe/H] $\simeq -1.3 \pm 0.2$.
We adopt this CMD-based value for the subsequent discussion.

Given the broad distribution of stars across the HB, it is also likely 
that Pal 2 contains RR Lyrae stars.  Although our original observations were
not extensive enough or well enough spaced to search for variables, we
note that the CMD of Fig.~7 has a dozen or more
candidate stars in the appropriate color range (near $(V-I) \sim 2$, taking
into account the foreground reddening discussed below).  
Finding and measuring any RR Lyraes in this cluster would be well worthwhile,
since it would help strengthen the reddening
and distance estimates considerably (see below).

\subsection{Reddening and Distance}

The {\it average} foreground reddening of Pal 2 can be estimated photometrically
through various features of the CMD compared with other ``standard'' clusters, or by
integrated colors.  We discuss each of these in turn.

\noindent {\it Integrated Colors:}  The intrinsic color of a globular
cluster is a well known but weakly sensitive function of metallicity
(\eg\ \markcite{rac73}Racine 1973; \markcite{ree88}Reed \etal~1988).
If the foreground absorption is already known to be large
(as it is here), and a reasonable guess can be made at the metallicity, then
the integrated color of the cluster is a very useful indicator of its reddening.
For the three comparison clusters listed above (NGC 1261, 1851, 6229) the mean
integrated colors are $(B-V)_0 \simeq 0.75 \pm 0.05, (V-I)_0 \simeq 0.95 \pm 0.05$.
In $(B-V)$ the listed color of Pal 2 is 2.08 (\markcite{har96}Harris 1996), giving
$E(B-V) = 1.33 \pm 0.07$.  

No published $(V-I)$ color for Pal 2 is available, but we generated one from our
new photometric data.  Directly adding all the stars from our cleaned CMD (Fig.~7), 
we obtain $(V-I)(int) = 2.65 \pm 0.05$ (the correction for faint main-sequence
stars below $V = 24$ is $\delta(V-I) \lesssim 0.03$ mag and is 
negligible for our purposes).  Subtracting the intrinsic color listed above, we
then obtain $E(B-V) = 0.75 E(V-I) = 1.28 \pm 0.07$.

\noindent {\it Horizontal Branch:}  The blue edge of the RR Lyrae gap, or
equivalently the reddest point on the blue HB, is insensitive to metallicity
and provides a useful reddening indicator for metal-poor clusters
(\eg\ \markcite{san69}Sandage 1969).  A {\it very rough} estimate of the 
position of the blue edge, from the color at which the HB starts to turn downward,
is $(V-I)_{be} = 2.0 \pm 0.2$.  The intrinsic color is $(V-I)_{0,be} = 0.25 \pm 0.05$
(\eg\ \markcite{wal94}Walker 1994), from which $E(B-V) = 1.3 \pm 0.2$.

\noindent {\it CMD Fitting:}  Fiducial sequences of clusters of similar metallicity
can be matched to the Pal 2 CMD in order to estimate both its reddening and
relative distance modulus.  In Fig. \ref{fig7}, we show 
mean lines for M3 (\markcite{joh97}Johnson \& Bolte
1997) and NGC 1851 (\markcite{dac90}Da Costa \& Armandroff 1990) 
superimposed on the CMD.  M3 is at slightly lower metallicity and thus is
placed just to the blue side of the Pal 2 RGB.  NGC 1851 should have nearly 
the same metallicity as Pal 2 (see above) and its RGB is matched to the center
of the RGB of Pal 2. The $\sim 0.05-$mag offset in color that we adopt here
between M3 and NGC 1851 is the
predicted amount for their metallicity difference, according to the calibrations
of \markcite{dac90} Da Costa \& Armandroff (1990).  
The fiducial sequence fit required a vertical shift of $\Delta(m-M)_V = 6.1$ mag 
relative to M3 -- determined mainly by the match to the HB stars -- 
and a reddening of $E(V-I) = 1.60 \pm 0.05$, thus $E(B-V) = 1.20 \pm 0.04$.  

Averaging all the reddening estimates listed
above, we obtain a final value of $E(B-V) = 1.24$.  The formal uncertainty
of the mean is only $\pm 0.03$ mag because of the close mutual agreement
of the individual estimates; however, the true uncertainty is likely to
be nearer $\pm0.07$ ($\pm 0.05$ from the individual estimates listed above,
coupled with an estimated $\sim 0.05-$mag combined uncertainty in the calibration
slopes and zero points when extended to these very red stars).  We emphasize again
that the mean reddening quoted here
should be taken as applying to the overall distribution
of stars near the center of the cluster, and not to any particular spot (which
could differ by more than $\pm 0.1$ mag due to differential reddening).

For the distance modulus, we use the HB level for Pal 2 of $V(HB) = 21.65 \pm 0.2$,
determined from the red HB clump and the stars near the (estimated) RR Lyrae blue edge.
With an assumed distance scale of $M_V(HB) = 0.7$ for moderately low metallicity
(\eg\ \markcite{van96}VandenBerg \etal~1996; \markcite{lay96}Layden \etal~1996),
we obtain $(m-M)_V = 20.95 \pm 0.2$ and hence an intrinsic distance modulus
$(m-M)_0 = 17.11 \pm 0.30$.  Pal 2 is then $(26 \pm 4$) kpc from the Sun
or $\sim 34$ kpc from the Galactic center.  Its approximate location in the $XZ$
Galactic coordinate plane is shown in Figure \ref{fig8}.  Only a 
handful of other clusters (NGC 2419, Palomar 3, 4, 14, AM-1, Eridanus, all of which
lie more than 80 kpc from the Galactic center) fall beyond the borders of this plot.

\section{Structural and Orbital Parameters}

The good seeing quality of our images, as noted above, permitted photometry
to be successfully carried in very close to the cluster center.  Inspection of
the DAOPHOT-generated images in which all measured stars had been subtracted,
as well as our artificial-star tests,
showed that for the bright stars in the CMD (the HB level and above), 
the completeness of detection was essentially 100\% for all radii larger
than $R \simeq 6''$.  Using the bright stars, we could therefore construct
radial profiles from direct starcounts and estimate the structural parameters
of the cluster.  To eliminate as much of the background
as possible and to avoid any problems with radially dependent photometric
incompleteness, we used only stars brighter than $V = 22$, and did not use
the northwest quadrant which is clearly hampered by heavier reddening (see 
Figs.~5 and 6).  The raw starcounts are listed in Table 1, where $R$ is the
mean radius in arcminutes of each annulus, defined as $R = (R_{min}R_{max})^{1/2}$;
$n$ is the number of stars brighter than $V = 22$ in the annulus;
and the area (given in arcmin$^2$) is the fraction of the complete
annulus that falls within the borders of the field.  Finally $\sigma$ is
the mean number density of stars within the annulus, in objects per arcmin$^2$,
with its Poisson uncertainty ($\pm n^{1/2}$) in the final column.
The resulting profile is shown graphically in Figure \ref{fig9}.

The starcounts are still declining slightly but noticeably past $R \gtsim 3'$.
Recognizing that the patchy reddening across the field makes the true
background level hard to define, we adopted $\sigma_b = (3.0\pm1.0)$
arcmin$^{-2}$ from the minimum of the outermost four radial bins, and 
subtracted this level from each point to define the cluster profile.
After subtraction of $\sigma_b$, the profile was fitted numerically by
standard isotropic \markcite{kin66}King (1966) models to deduce the
central concentration and scale radii of the cluster.
Although our starcounts do not penetrate in to the very center, we find no
evidence for a `collapsed core'; the basic King-type model provides an
adequate fit to the observations at all radii.
The best-fitting curve gives a core radius $r_c = 0\farcm 24 \pm 0\farcm 06$, 
half-mass radius $r_h = 0\farcm 67 \pm 0\farcm 09$, and central concentration 
$c = 1.45 \pm 0.26$.  This model fit is shown in Figure \ref{fig10}.
The uncertainties quoted above include both the statistical uncertainty in the
fit and the added effect of the $\pm 1.0$ arcmin$^{-2}$ uncertainty in the
adopted background level.  For comparison, the previous catalog values
(\markcite{tra95}Trager \etal\ 1995) are $c = 1.91$ and $r_c = 0\farcm 15$,
noticeably but not extremely different from our measurements.
We regard our new estimates as preferable, since 
the starcounts shown here define the cluster
profile over a much larger run of radius than did the earlier material
(see Fig.~2 of \markcite{tra95} Trager \etal), and by selection
of stars from the CMD we have
been able to reduce the background to very low levels.

The luminosity of the cluster, and thus its total mass, can be estimated
from its integrated $V$ magnitude and our distance modulus quoted above.
The catalog value (\markcite{har96}Harris 1996) is $V_t = 13.04$, which
compares well with our own direct measurement of $V_t \simeq 13.05 \pm 0.1$ 
within a large aperture of radius 300 px ($63''$) centered on the 
cluster.  Subtracting $(m-M)_V = 20.95 \pm 0.2$ then yields an
integrated luminosity $M_V^t = -7.9 \pm 0.25$, which is $\sim 0.4$ mag
brighter than the mean luminosity of the globular clusters in the Milky Way
(\markcite{har91}Harris 1991).  For a typical mass-to-light ratio $(M/L)_V \sim 2.5$ 
(\markcite{mey97}Meylan \& Heggie 1997; \markcite{pry93}Pryor \& Meylan 1993; 
\markcite{man91}Mandushev \etal~1991)
the total cluster mass is then $M_t \simeq 2.8 \times 10^5 M_{\odot}$.
Pal 2 is one of the brightest and most massive clusters in the outer halo,
clearly exceeded only by NGC 2419 (at $M_V^t \simeq -9.5$).

The radial velocity of Pal 2, based on a single uncertain measurement
(\markcite{har78}Hartwick \& Sargent 1978; \markcite{web81}Webbink 1981), 
is $v_r = -133 \pm 57$ km/s, or $v_{LSR} = -142$ km/s corrected to the
solar Local Standard of Rest (\markcite{har96}Harris 1996).
Since our line of sight toward Pal 2 is very close to the Galactic anticenter,
$v_{LSR}$ very nearly represents 
its radial velocity relative to the Galactic center; thus, 
its apogalacticon must be much further out than its present
location, and it must now be coming in towards perigalacticon.
A very rough estimate of the perigalactic distance $R_p$ can
be made from the structural parameters listed above:
our measured core and half-mass radii correspond 
to $r_c = (1.8 \pm 0.5)$ pc and $r_h = (5.1 \pm 0.9)$ pc.
These values fall in the midrange for scale sizes of Milky Way globular
clusters (\eg\ \markcite{van95}van den Bergh 1995), and are much smaller
than the scale radii for (\eg) the half-dozen outermost-halo clusters mentioned
above.  The well known
correlation of half-mass radius with either Galactocentric distance $R_{gc}$
or perigalactic distance $R_p$ 
(see Fig.~6 of \markcite{van95}van den Bergh 1995) then suggests that $R_p$ 
for Pal 2 is significantly smaller than its present distance:
reading from van den Bergh's empirically deduced correlation between 
$R_p$ and $r_h$, we estimate that $R_p$ should be in the range of 5 to 10 kpc,
\ie\ somewhere near the Solar circle and about one-third to one-quarter
its present Galactocentric distance.  Knowing that it is well between both
apogalacticon and perigalacticon, we can take its present distance $R_{gc}$ as
a reasonable estimate of the orbital semimajor axis $A$, defined as
in \markcite{inn83}Innanen \etal\ (1983; here $A$ is the radius of a circular
orbit with the same orbital energy as the cluster).  If this argument
is correct, then the cluster's orbital 
eccentricity $e \equiv (1-R_p/A)$ must be quite high ($e \gtsim 0.7$) and
its transverse velocity or proper motion correspondingly small.
Other methods for estimating the orbital parameters depend on 
knowledge of the cluster's {\it tidal} radius (\cf
\markcite{inn83} Innanen \etal~1983),
which unfortunately we do not know reliably (and which will
be extremely hard to measure, given the patchy reddening
across the field and its effect on the starcounts).

\section{Summary}

	New CFHT photometry in $V$ and $I$ has been used to measure the color-magnitude
diagram of Palomar 2 to a limiting magnitude just short of its main-sequence turnoff.  
Its CMD morphology is that of an intermediate-metallicity cluster near [Fe/H] $\sim -1.3$,
similar to those of some other globular clusters in the mid- to outer-halo.
Correcting for its very large foreground reddening $E(B-V) \simeq 1.24$, we find that
it is now about 34 kpc from the Galactic center; 
very rough arguments based on its structural parameters and radial velocity 
suggest that it is now moving in towards perigalacticon on a highly elliptical orbit.

	Additional studies of Pal 2 would be well worthwhile, but 
will always have to cope with its frustratingly high reddening and differential 
reddening.  One obvious way to alleviate these problems would be to carry out 
a study of the CMD in the near infrared $(JHK)$, which should
be able to sharpen up the definition of the principal sequences.  Another very 
helpful new observation would be a direct spectroscopic determination of its metallicity
and radial velocity, both of which are poorly known at present.
Finally, we predict on the basis of the CMD morphology that Pal 2 should
have a dozen or more RR Lyrae variables; a survey of these would 
considerably help to refine the reddening and distance estimates.

\acknowledgements
We are happy to thank Dean McLaughlin for helping us with the King-model
fits.  This work was supported in part by the Natural Sciences and Engineering 
Research Council of Canada, through an operating grant to WEH and travel
grants to the CFHT.

\begin{deluxetable}{crcrr}
\tablewidth{300pt}
\tablecaption{Starcounts for Bright Stars in Palomar 2 \label{tab1}}
\tablecolumns{5}
\tablehead{
\colhead{$R'$} & \colhead{$n$} & \colhead{area} &
\colhead{$\sigma$} & \colhead{$\pm$} 
}
\startdata
 0.096 &  7 &  0.008& 885.93 & 334.85  \nl
 0.121 & 20 &  0.020& 987.68 & 220.85  \nl
 0.157 &  9 &  0.026& 347.21 & 115.74  \nl
 0.192 & 18 &  0.032& 567.33 & 133.72  \nl
 0.242 & 41 &  0.081& 508.34 &  79.39  \nl
 0.313 & 37 &  0.104& 356.43 &  58.60  \nl
 0.429 & 79 &  0.361& 219.05 &  24.65  \nl
 0.606 & 69 &  0.505& 136.61 &  16.45  \nl
 0.783 & 53 &  0.649&  81.65 &  11.22  \nl
 0.959 & 25 &  0.794&  31.50 &   6.30  \nl
 1.212 & 43 &  2.020&  21.28 &   3.25  \nl
 1.565 & 47 &  2.597&  18.10 &   2.64  \nl
 1.917 & 51 &  3.175&  16.07 &   2.25  \nl
 2.268 & 30 &  3.752&   8.00 &   1.46  \nl
 2.619 & 32 &  4.329&   7.39 &   1.31  \nl
 2.970 & 17 &  4.847&   3.51 &   0.85  \nl
 3.320 & 20 &  4.450&   4.49 &   1.01  \nl
 3.671 & 13 &  4.002&   3.25 &   0.90  \nl
 4.021 & 18 &  3.007&   5.99 &   1.41  \nl
\enddata
\end{deluxetable}

\newpage
\begin{center}
Figure Captions
\end{center}

\figcaption[Harris.fig1a.ps]{(a) A 300-second exposure of the Palomar 2 field
in $V$, taken at the CFHT prime focus.  North is up and East is to the left.
The field size is $7\farcm 2$ across.  
(b) The central part of the Pal 2 field,
extracted from the previous image.  The field size shown here is $2\farcm 1$ across.
\label{fig1}}

\figcaption[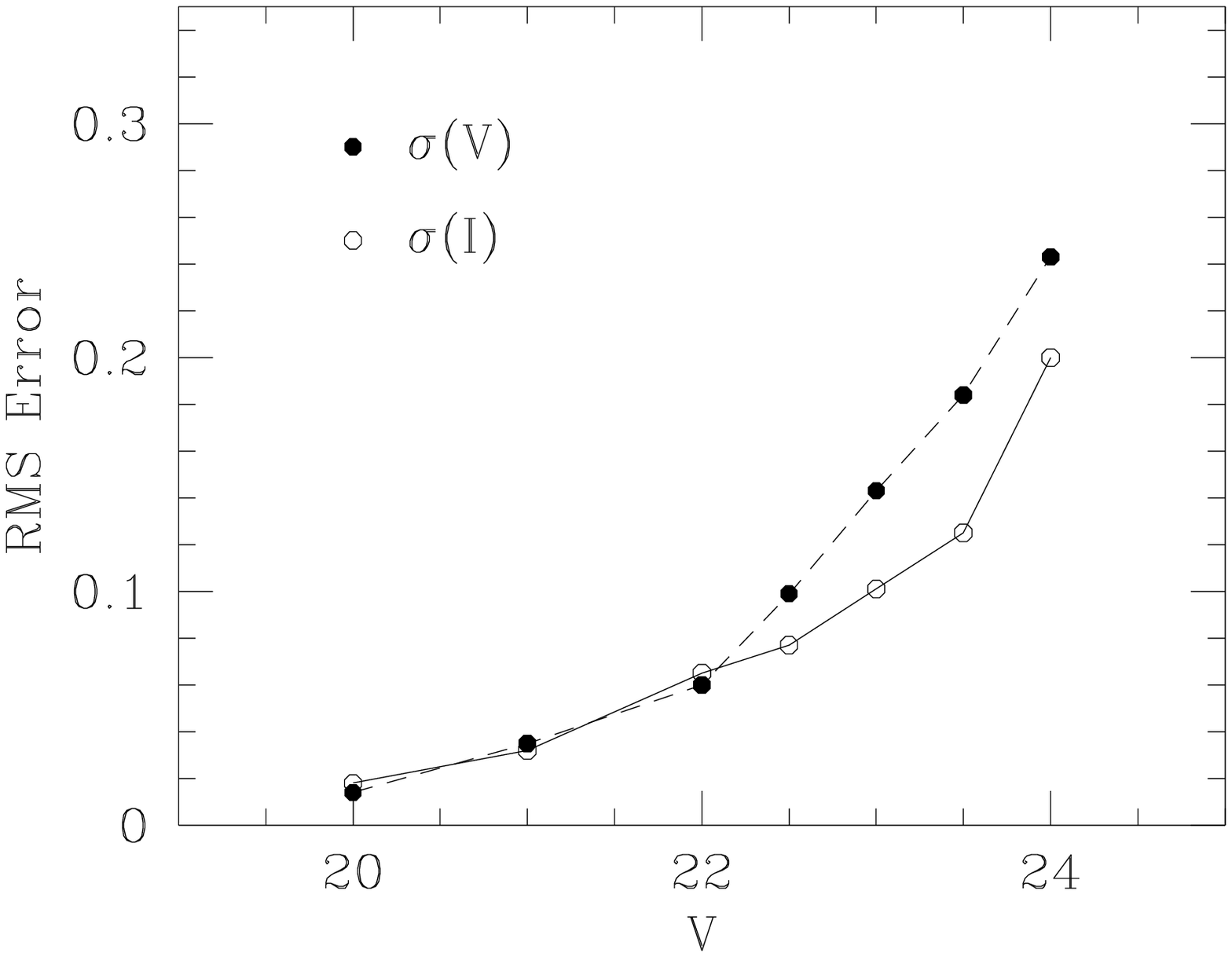]{Internal measurement uncertainties of the photometry,
derived from the artificial-star tests described in the text.  Here $\sigma(V),
\sigma(I)$ are the rms errors for test stars at the $V$ magnitude plotted,
and at $(V-I) = 2.5$ (the color of the cluster giants).
\label{fig2}}

\figcaption[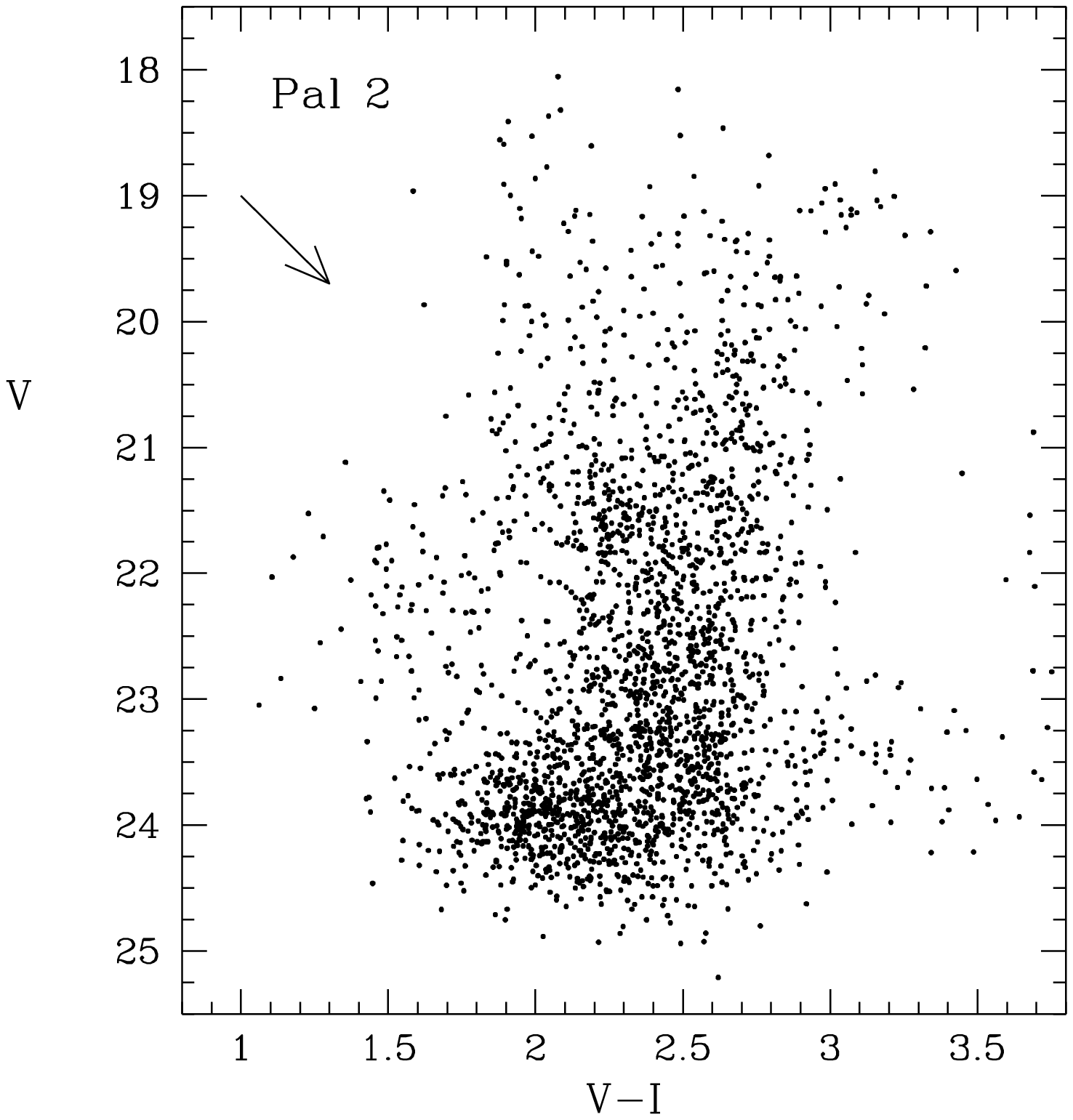]{Color-magnitude diagram for all measured stars in the
Pal 2 field.  A reddening vector of $\Delta E(V-I) = 0.25$ is shown at upper left.
\label{fig3}}

\figcaption[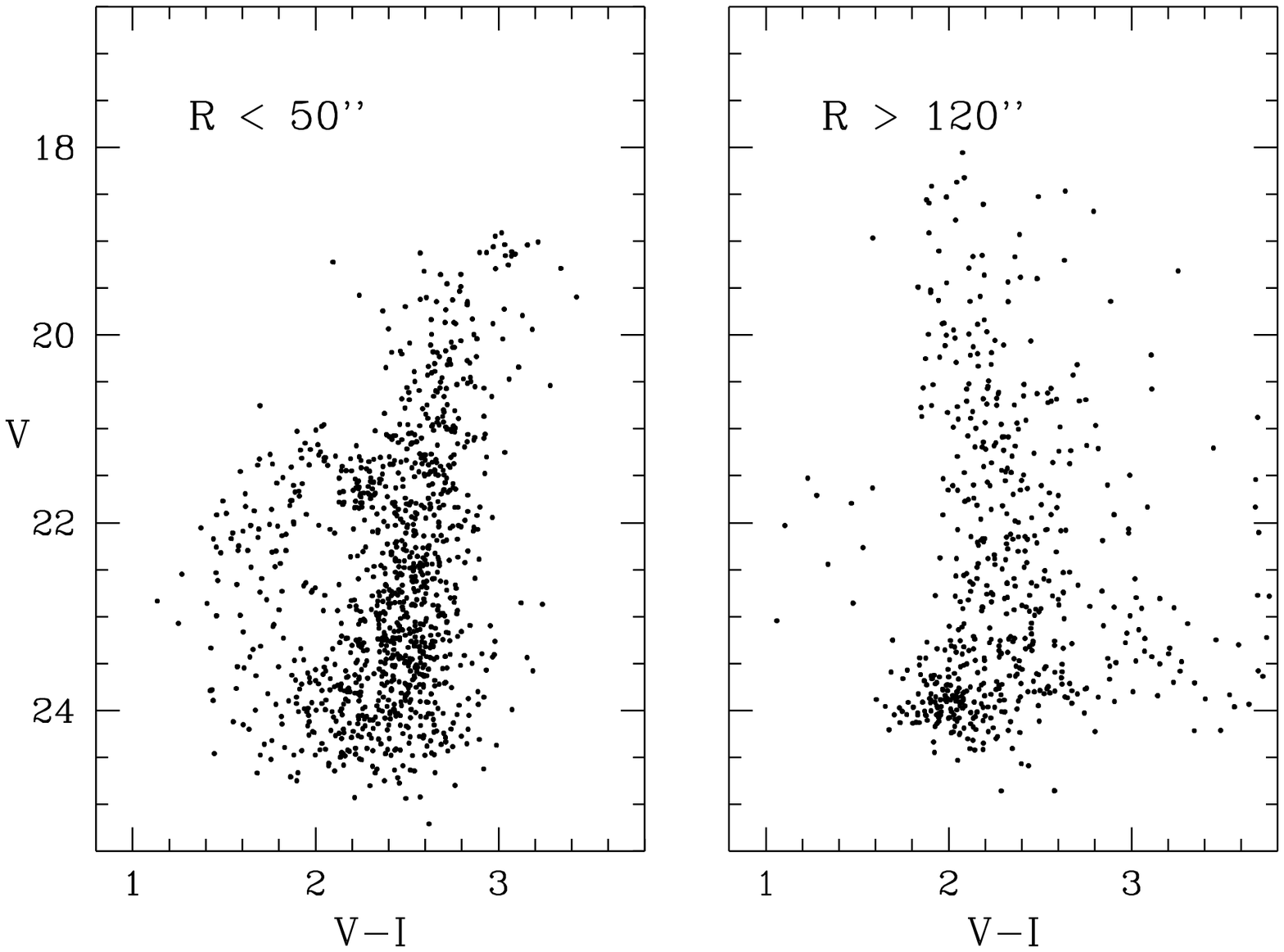]{(a) Color-magnitude diagram for an ``inner'' sample
of stars centered on the cluster (stars within radius $R < 50''$).  (b) CMD for 
an ``outer'' sample ($R > 120''$), dominated by the background field population.
\label{fig4}}

\figcaption[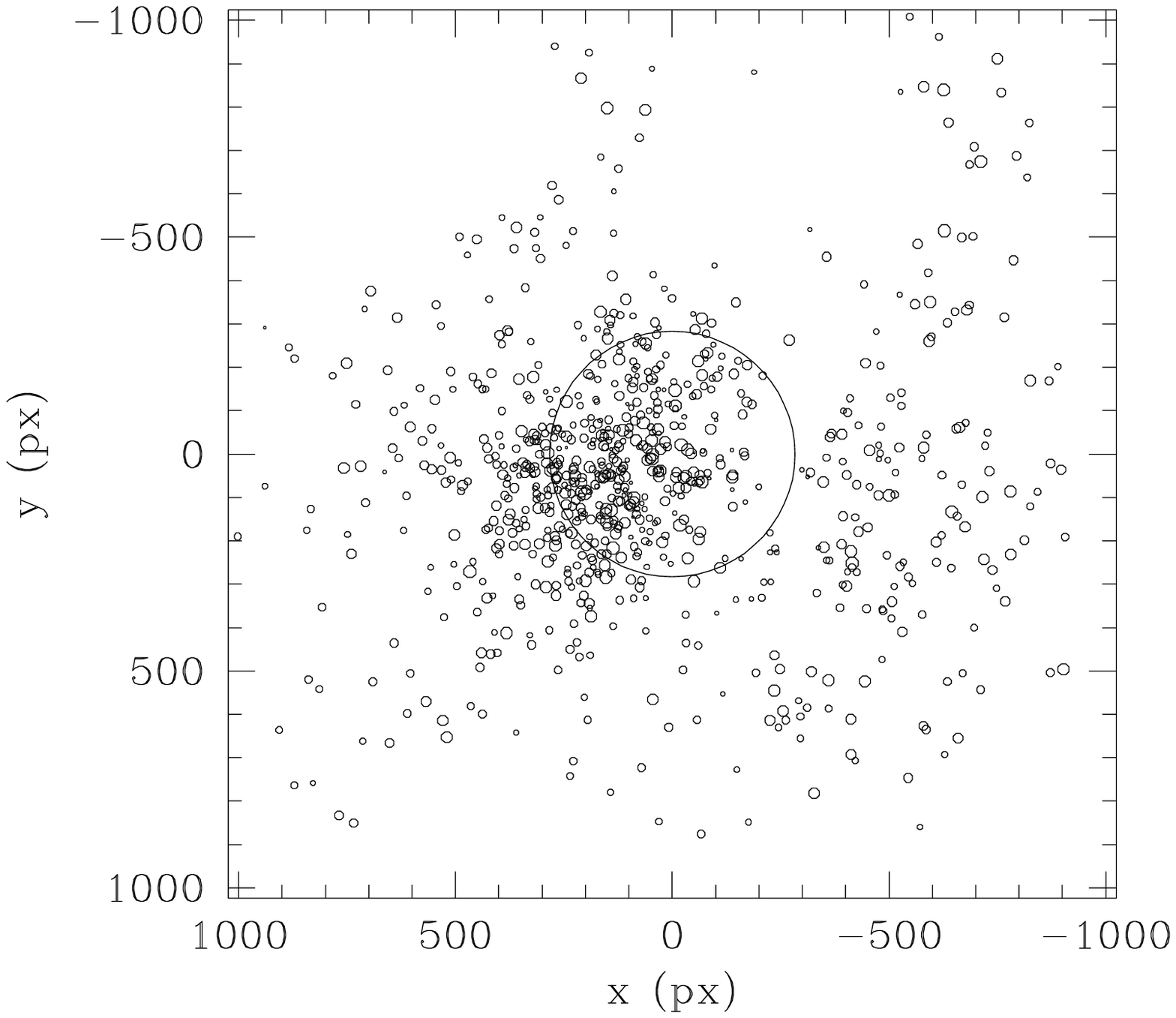]{Spatial distribution of the faint blue stars 
in the cluster field (those with $V > 23$, $(V-I) < 2.4$).  Individual
stars are coded by size, with brighter stars plotted in larger symbols.
Many or most of these should be subgiant and turnoff stars from the cluster, and provide a
useful map of the differential reddening across the field; see \S3 of the
text.  Note the N/S lane of higher than average reddening running just to
the right of the cluster, as well as the smaller region SE of cluster center
which has a higher proportion of faint stars.  \label{fig5}}

\figcaption[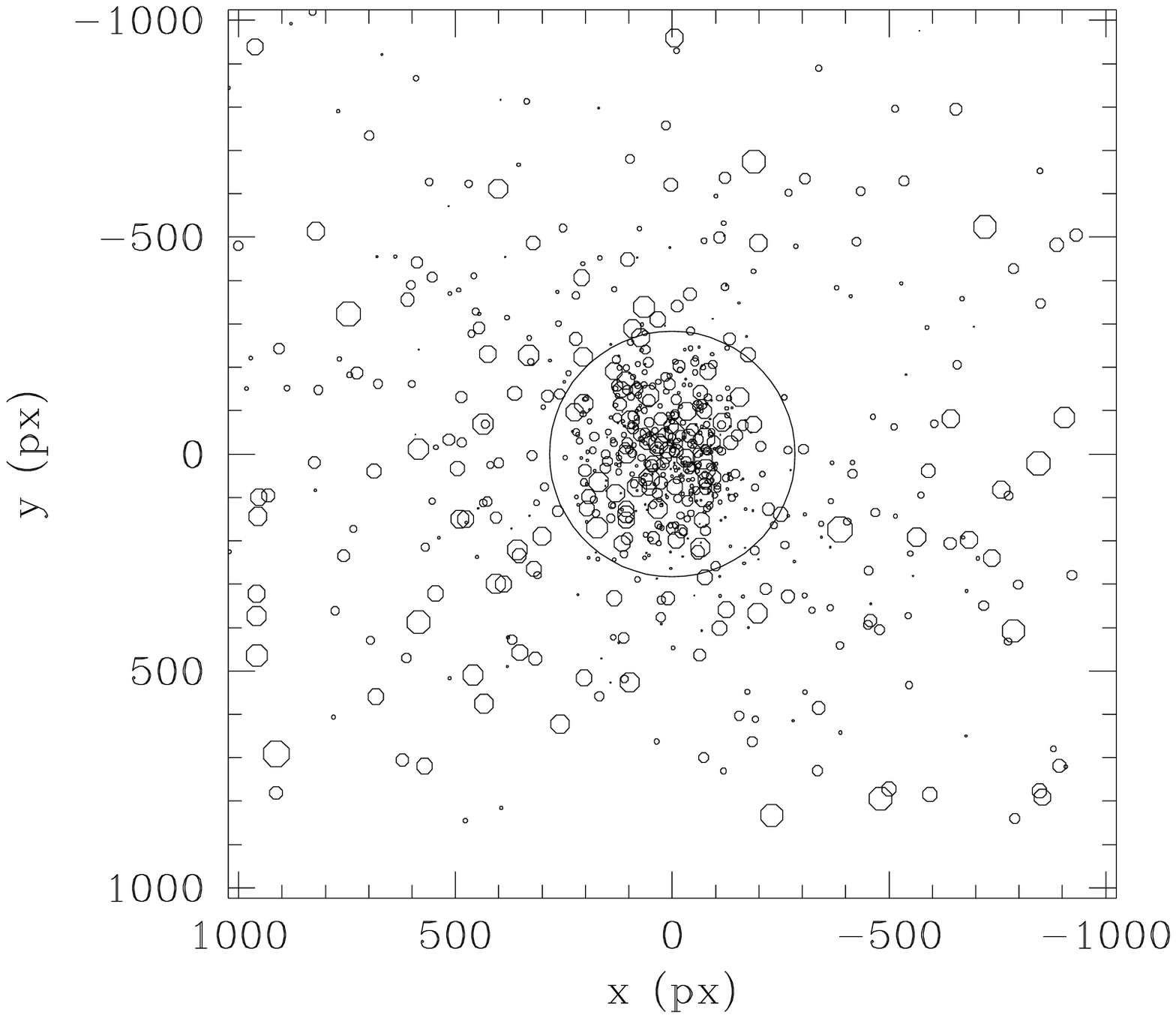]{Spatial distribution of the bright stars
($V < 22$) in the Pal 2 field.  Larger symbols correspond to brighter stars.
Except for the more heavily reddened NW quadrant (upper right), 
the distribution is much more even than in the previous figure.
\label{fig6}}

\figcaption[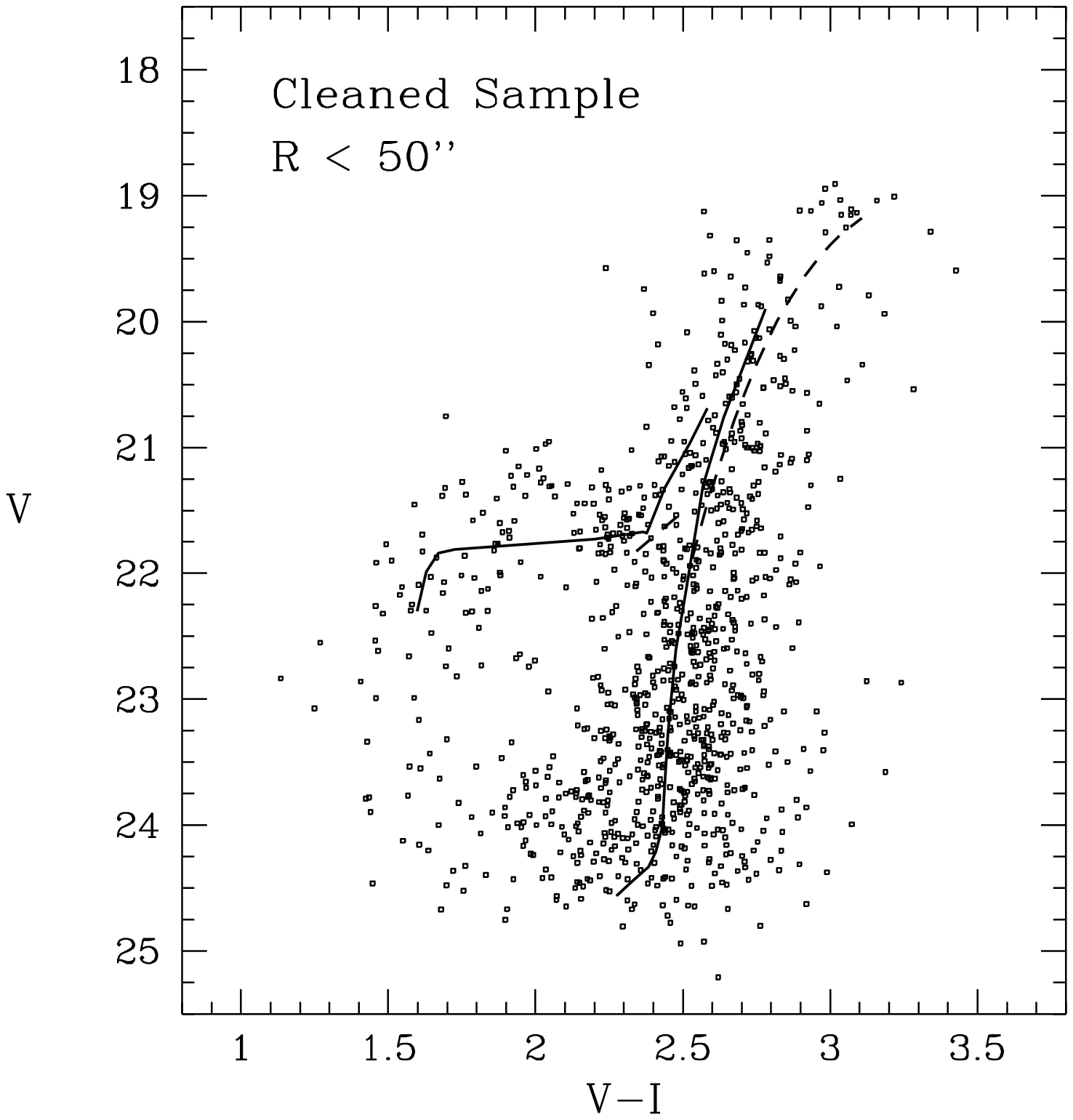]{CMD in $(I,V-I)$ for the central region in
Fig.~4a.  Field-star contamination has been statistically removed as
described in \S3.  The {\it solid lines} superimposed on the Pal 2 data
are the fiducial sequences for the standard cluster M3, shifted to the
red by $\Delta(V-I) = 1.60$ (see \S3.2 of the text).  The {\it dashed line}
is the fiducial RGB line for NGC 1851, a cluster which has the same CMD
morphology as Pal 2.  \label{fig7}}

\figcaption[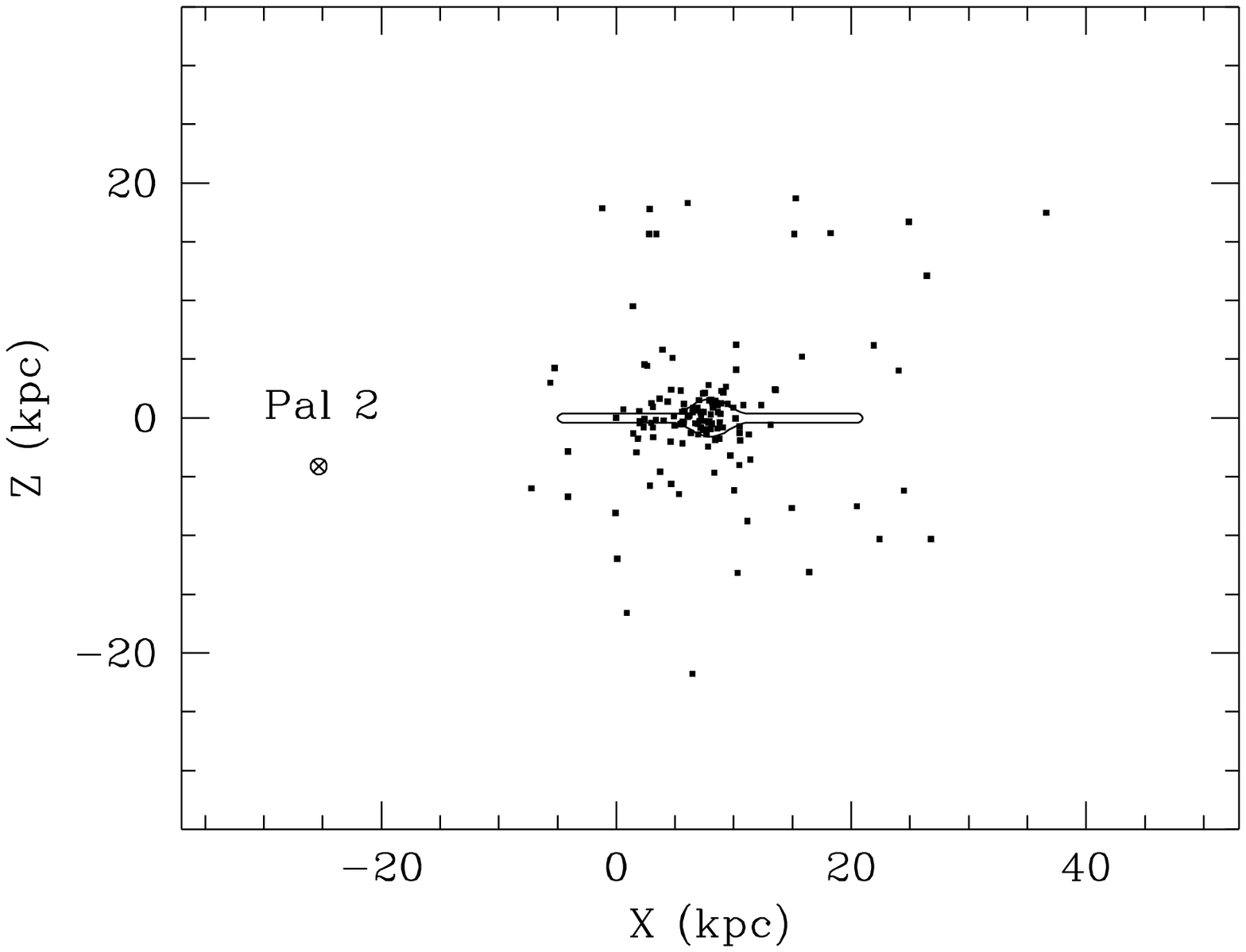]{Spatial distribution of Milky Way globular
clusters, in the XZ Galactic coordinate plane.  In this plane the Sun is
at (0,0) and the Galactic center at (8,0) kpc.  Data are taken from
the catalog of \protect\markcite{har96} Harris (1996).  The only known halo clusters
outside this plot are the six extreme outer-halo objects NGC 2419, Pal 3, 4, 14,
Eridanus, and AM-1.  \label{fig8}}

\figcaption[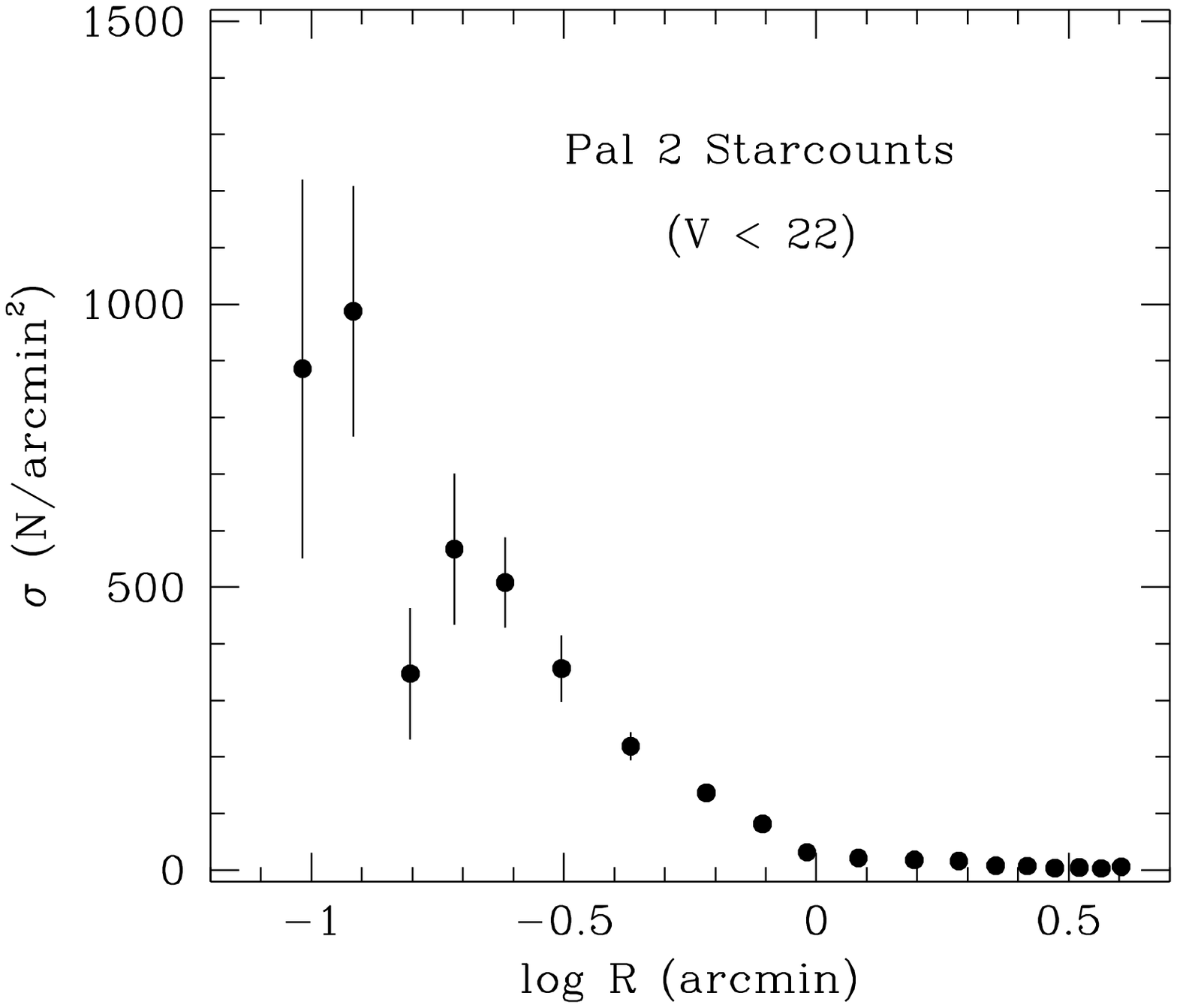]{Radial distribution of the bright stars
$(V < 22)$ in the Pal 2 field.  Here $\sigma$ is the number of stars
per unit area, plotted against radius $R$ in arcmin.
\label{fig9}}

\figcaption[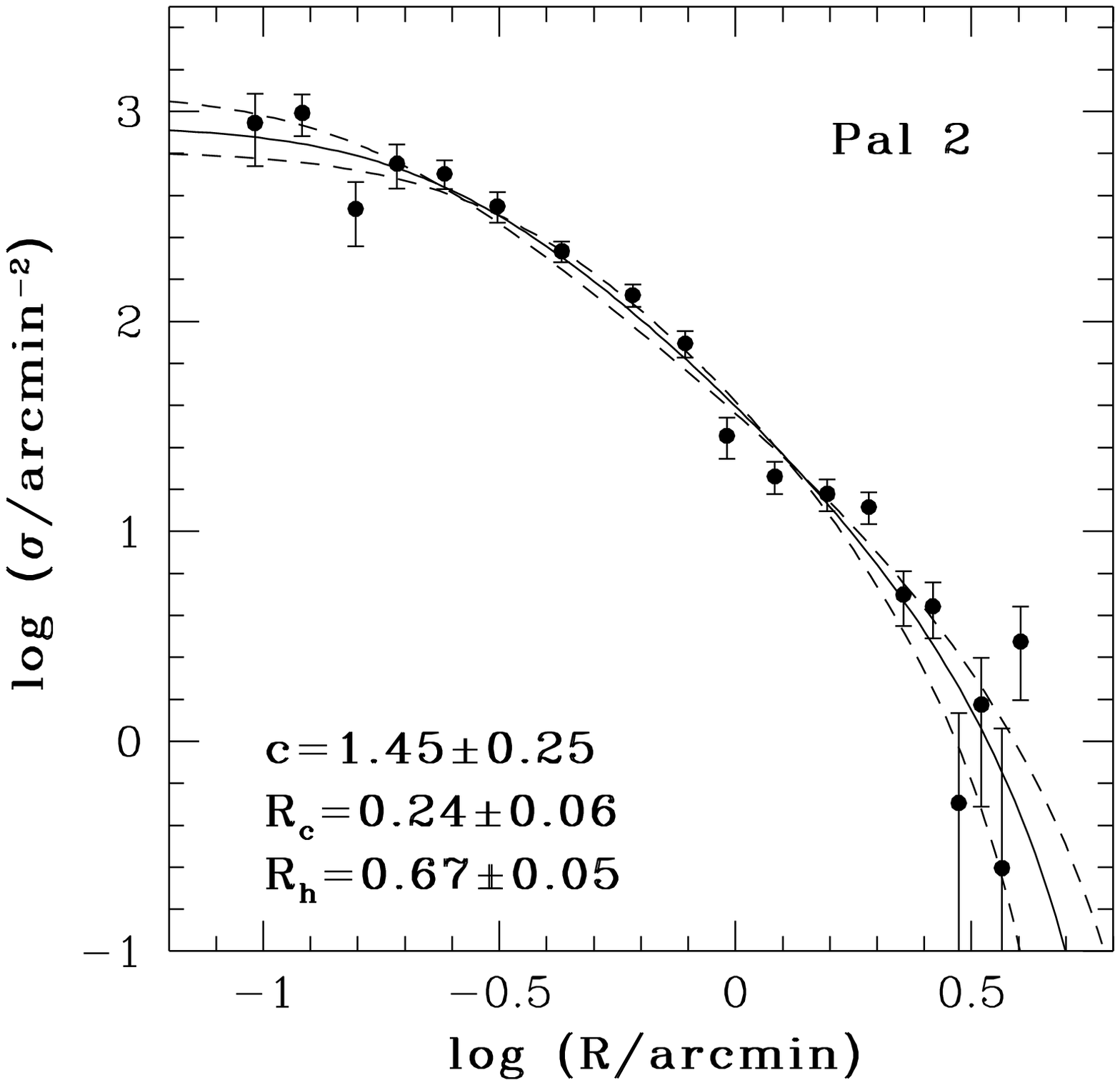]{Model fits to the starcounts, after subtraction
of the background count level $\sigma_b$ (see \S4 of the text).
The solid line represents the best-fitting King (1966) model with
core radius $r_c$ and central concentration $c$, while the
dashed lines indicate the range of solutions within the quoted errors.
The uncertainties given in the figure include only the internal
uncertainty of the fit and do not include the additional uncertainty in
$\sigma_b$; see text.  \label{fig10}}

\clearpage
\begin{figure}
\epsscale{1.0}
\plotone{Harris.fig2.ps}
\end{figure}
 
\clearpage
\begin{figure}
\epsscale{1.0}
\plotone{Harris.fig3.ps}
\end{figure}
 
\clearpage
\begin{figure}
\epsscale{1.0}
\plotone{Harris.fig4.ps}
\end{figure}
 
\clearpage
\begin{figure}
\epsscale{1.0}
\plotone{Harris.fig5.ps}
\end{figure}
 
\clearpage
\begin{figure}
\epsscale{1.0}
\plotone{Harris.fig6.ps}
\end{figure}
 
\clearpage
\begin{figure}
\epsscale{1.0}
\plotone{Harris.fig7.ps}
\end{figure}
 
\clearpage
\begin{figure}
\epsscale{1.0}
\plotone{Harris.fig8.ps}
\end{figure}
 
\clearpage
\begin{figure}
\epsscale{1.0}
\plotone{Harris.fig9.ps}
\end{figure}
 
\clearpage
\begin{figure}
\epsscale{1.0}
\plotone{Harris.fig10.ps}
\end{figure}
 
\end{document}